\documentclass{aipproc}
\layoutstyle{6x9}
\begin{document}

%\title{ Study of the strong  $\Sigma_c\to \Lambda_c\, \pi$, 
%$\Sigma_c^{*}\to \Lambda_c\, \pi$ and
%$\Xi_c^{*}\to \Xi_c\, \pi$ decays in a nonrelativistic quark model}
\title{Strong one-pion decay of $\Sigma_c$, $\Sigma_c^*$ and $\Xi_c^*$}
\author{C. Albertus}{address={Departamento de F\'{\i}sica At\'omica,
    Molecular y Nuclear, Universidad de
Granada, E-18071 Granada, Spain.}}
\author{ E. Hern\'andez}{address={Grupo de F\'\i sica Nuclear, Departamento de
F\'\i sica Fundamental e IUFFyM, 
Facultad de Ciencias, E-37008 Salamanca, Spain.}}  
\author {J. Nieves}{address={Departamento de F\'{\i}sica At\'omica,
    Molecular y Nuclear, Universidad de
Granada, E-18071 Granada, Spain.}}
\author{ J. M. Verde-Velasco}{address={Grupo de F\'\i sica Nuclear, Departamento de
F\'\i sica Fundamental  e IUFFyM, 
Facultad de Ciencias, E-37008 Salamanca, Spain.}} 


%\maketitle

\begin{abstract}
Working in the framework of a nonrelativistic quark model we  
evaluate the  widths for the strong one-pion decays
$\Sigma_c\to\Lambda_c\,\pi$, $\Sigma_c^*\to\Lambda_c\,\pi$ and 
$\Xi_c^{*}\to\Xi_c\,\pi$. We take  advantage of the constraints imposed by
heavy quark symmetry to  solve the three-body problem by means of a simple
variational ansatz. 
We use partial conservation of the axial  current hypothesis  to get the strong
vertices from weak axial current matrix
elements.  Our results are in good agreement
with experimental data.
\end{abstract}

\maketitle
%We evaluate the widths for the charmed baryon
%$\Sigma_c\to\Lambda_c\,\pi$, $\Sigma_c^*\to\Lambda_c\,\pi$ and 
%$\Xi_c^{*}\to\Xi_c\,\pi$ strong decays. 

%The calculations have been done within a
%nonrelativistic constituent quark model. To the best of our knowledge 
%this is the first time that a fully dynamical calculation of these 
%observables has been done within a nonrelativistic approach, this has
%been possible with the use of wave functions 
%constrained by heavy quark symmetry and 
%that

In this contribution we present results  for the  widths of the strong
one-pion decay processes
$\Sigma_c\to\Lambda_c\,\pi$, $\Sigma_c^*\to\Lambda_c\,\pi$ and 
$\Xi_c^{*}\to\Xi_c\,\pi$, obtained within a
nonrelativistic quark model. To the best of our
knowledge this is the first fully dynamical calculation of these
observables done within a nonrelativistic approach.
We use heavy quark symmetry constraints  
on baryons with a heavy quark to solve the three-body problem by means of a 
simple variational ansazt. 
The orbital wave functions thus obtained are simple and manageable. Their
functional form and the corresponding variational parameters are given 
in Ref.~\cite{albertus04}. In order to check the dependence of the results
on the interquark interaction we have used five different quark-quark 
potentials that we took from Refs.~\cite{bhaduri81,silvestre96}. All the potentials include a confining
term plus Coulomb and hyperfine terms coming from one-gluon exchange, while they
differ in the  power of the confining term and/or in the different
 regularization of the singular behaviour of the one-gluon terms at
the origin.
The pion emission amplitude is evaluated in a one-quark pion emission model
(spectator approximation) in which we use partial conservation of the axial
current to determine the strong couplings through the evaluation of axial
matrix elements.
Due to the limitations  of space, we shall focus on
the presentation of the results and their comparison with experimental data.
Nevertheless our tables also show other theoretical results obtained  
using the  constituent
quark model (CQM),  heavy hadron chiral perturbation theory
(HHCPT), and   relativistic quark models like the light-front quark model
(LFQM) and the relativistic three-quark model (RTQM). 
Full details on our calculation are given in Ref.~\cite{nuevo}.

%\section{Results}

\begin{table}[h!!!!!!!!]
%\begin{center}
\begin{tabular}{l c c c}
%\hline
   &
 \hspace{.25cm}$\Gamma(\Sigma_c^{++}\to\Lambda_c^+\pi^+)$ \hspace{.25cm} &
 \hspace{.25cm} $\Gamma(\Sigma_c^{+}\to\Lambda_c^+\pi^0)$ \hspace{.25cm} &
 \hspace{.25cm}$\Gamma(\Sigma_c^{0}\to\Lambda_c^+\pi^-)$ \hspace{.25cm} 
 \\
 &[MeV]&[MeV]&[MeV]\\
\hline
%This work AL1 & $21.84\pm0.08$ &$2.43\pm 0.02$ &$2.81\pm 0.02$ &$2.39\pm 0.02$\\
%This work AL2 & $21.88\pm0.08$ &$2.44\pm 0.02$ &$2.84\pm 0.02$ &$2.40\pm 0.02$\\
%This work AP1 & $21.44\pm0.09$ &$2.35\pm 0.02$ &$2.71\pm 0.02$ &$2.31\pm 0.02$\\
%This work AP2 & $21.41\pm0.08$ &$2.34\pm 0.02$ &$2.70\pm 0.02$ &$2.30\pm 0.02$\\
%This work BHAD &$22.06\pm0.08$ &$2.48\pm 0.02$ &$2.87\pm 0.02$ &$2.44\pm 0.02$\\

%&  & &\\
This work 
%avg. 
 &$2.41\pm0.07\pm0.02$ &$2.79\pm0.08\pm0.02$ 
&$2.37\pm0.07\pm0.02$\\
\hline
Experiment   & $2.3\pm0.2\pm 0.3$~\cite{cleo02} & $< 4.6$ (C.L.=90\%)~\cite{cleo01} & 
$2.5\pm0.2\pm0.3$~\cite{cleo02} \\
  &$2.05^{+0.41}_{-0.38}\pm0.38$~\cite{focus02} & &
 $1.55^{+0.41}_{-0.37}\pm0.38$~\cite{focus02}\\
 \hline
Theory  &  & &\\
CQM
&  $1.31\pm 0.04$~\cite{rosner95} &$1.31\pm 0.04$~\cite{rosner95}& 
$1.31\pm 0.04$~\cite{rosner95}\\
 & $2.025^{+1.134}_{-0.987}$~\cite{pirjol97} & &
 $1.939^{+1.114}_{-0.954}$~\cite{pirjol97} \\
HHCPT  & 2.47, 4.38~\cite{yan92}& 2.85,
5.06~\cite{yan92}&2.45, 4.35~\cite{yan92}\\
 &  2.5~\cite{huang95}&  3.2~\cite{huang95}& 2.4~\cite{huang95}\\
 &  &  &$1.94\pm0.57$~\cite{cheng97}\\
LFQM &  1.64 ~\cite{tawfiq98}&1.70 ~\cite{tawfiq98} & 1.57 ~\cite{tawfiq98}\\
RTQM & $2.85\pm0.19$~\cite{ivanov9899} &$3.63\pm0.27$~\cite{ivanov9899}
  & $2.65\pm0.19$~\cite{ivanov9899} \\
\hline
\end{tabular}
%\end{center}
\caption{\mbox{\hspace{-.2cm} 
Total decay widths for $\Gamma(\Sigma_c^{++}\to\Lambda_c^+\pi^+)$,
$\Gamma(\Sigma_c^{+}\to\Lambda_c^+\pi^0)$ and
$\Gamma(\Sigma_c^{0}\to\Lambda_c^+\pi^-)$.}
}
\label{tab:siglam}
%\vspace{-1cm}
\end{table}

Our results for the  $\Sigma_c$ one-pion decay widths  are given  in
Table  \ref{tab:siglam}. They include two different classes of errors:
the first one reflects the change of the results with the potential used, while
the second  is  purely numerical.  
Our results are in very good agreement with  experimental data by
CLEO ~\cite{cleo02,cleo01} and in a reasonable agreement with  data by
FOCUS~\cite{focus02}.

In Table~\ref{tab:sig*lam} we present  the results for the  $\Sigma^*_c$
one-pion decay widths.
Our central value for $\Gamma(\Sigma_c^{*\,++}\to\Lambda_c^+\pi^+)$ is above 
the central value of the
latest experimental data  by CLEO~\cite{cleo05}. We get results  within 
experimental errors
for the AP1 and AP2 potentials of Ref.~\cite{silvestre96}. 
For $\Gamma(\Sigma_c^{*\,+}\to\Lambda_c^+\pi^0)$ our  central value
is slightly above the upper experimental bound obtained  by 
CLEO ~\cite{cleo01}, while for  the AP1 and AP2 potentials we are below that
bound.
In the case of $\Gamma(\Sigma_c^{*\,0}\to\Lambda_c^+\pi^-)$ decay 
we agree with experiment.\\

\begin{table}[h!!!]
%\begin{center}
\begin{tabular}{l c c c}
%\hline
 & \hspace{.25cm}$\Gamma(\Sigma_c^{*\,++}\to\Lambda_c^+\pi^+)$ \hspace{.25cm}
 & \hspace{.25cm}$\Gamma(\Sigma_c^{*\,+}\to\Lambda_c^+\pi^0)$ \hspace{.25cm}
 & \hspace{.25cm}$\Gamma(\Sigma_c^{*\,0}\to\Lambda_c^+\pi^-)$ \hspace{.25cm}\\
 &[MeV]&[MeV]&[MeV]\\
\hline
%This work AL1 & $36.54\pm0.13$ &$17.85\pm0.12$&$17.63\pm0.13$ &$17.21\pm0.12$\\
%This work AL2 & $36.56\pm0.12$ &$17.86\pm0.12$&$17.65\pm0.12$ &$17.23\pm0.11$\\
%This work AP1 & $35.43\pm0.13$ &$16.78\pm0.12$&$16.58\pm0.12$ &$16.18\pm0.12$\\
%This work AP2 & $35.53\pm0.13$ &$16.87\pm0.12$&$16.67\pm0.12$ &$16.27\pm0.12$\\
%This work BHAD& $36.96\pm0.12$ &$18.26\pm0.12$&$18.04\pm0.12$ &$17.61\pm0.11$\\

% && &\\
This work 
%avg. 
&$17.52\pm0.74\pm0.12$&$17.31\pm0.73\pm0.12$ &
$16.90\pm0.71\pm0.12$\\
\hline
Experiment   & $14.1^{+1.6}_{-1.5}\pm 1.4$~\cite{cleo05}& $< 17$ (C.L.=90\%)~\cite{cleo01} & $16.6^{+1.9}_{-1.7}\pm1.4$~\cite{cleo05}\\
\hline
Theory &&&\\
CQM & 20~\cite{rosner95}&20~\cite{rosner95} &20~\cite{rosner95}\\
HHCPT    & 25~\cite{huang95}&25~\cite{huang95} &25~\cite{huang95}\\
LFQM    & 12.84~\cite{tawfiq98} & & 12.40~\cite{tawfiq98}\\
RTQM& $21.99\pm0.87$~\cite{ivanov9899} &
  & $21.21\pm0.81$~\cite{ivanov9899} \\
\hline
\end{tabular}
%\end{center}
\caption{\mbox{\hspace{-.3cm}  
Total decay widths for $\Gamma(\Sigma_c^{*\,++}\to\Lambda_c^+\pi^+)$,
$\Gamma(\Sigma_c^{*\,+}\to\Lambda_c^+\pi^0)$ and
$\Gamma(\Sigma_c^{*\,0}\to\Lambda_c^+\pi^-)$.} }
\label{tab:sig*lam}
\end{table}%\vspace{1cm}

Finally in Table \ref{tab:cas*cas} we present results for
partial and total  $\Xi^*_c$ one-pion decay widths. 
Our central value for $\Gamma(\Xi_c^{*\,+}\to\Xi_c^0\pi^++\Xi_c^+\pi^0)$ 
is slightly
above the experimental bound obtained by CLEO \cite{cleo96}. As before 
our results for the AP1 and AP2 potentials are below that  bound.
 For  
$\Gamma(\Xi_c^{*\,0}\to\Xi_c^+\pi^-+\Xi_c^0\pi^0)$ our result
is clearly smaller than the experimental upper bound  determined by CLEO~\cite{cleo95}. \\
\begin{table}[h!!!!!]
%\begin{center}
\begin{tabular}{lc c c c c}
%\hline 
 &\hspace{.01cm} $\Gamma(\Xi_c^{*\,+}\to\Xi_c^0\pi^+)$ \hspace{.01cm} 
 &\hspace{.01cm} $\Gamma(\Xi_c^{*\,+}\to\Xi_c^+\pi^0)$ \hspace{.01cm} 
 &\hspace{.01cm} $\Gamma(\Xi_c^{*\,0}\to\Xi_c^+\pi^-)$ \hspace{.01cm}
 &\hspace{.01cm} $\Gamma(\Xi_c^{*\,0}\to\Xi_c^0\pi^0)$ \hspace{.01cm} \\
 &[MeV] & [MeV]&[MeV]&[MeV]\\ 
\hline
This work  
%avg. 
&$1.84\pm0.06\pm0.01$&
$1.34\pm0.04\pm0.01$&$2.07\pm0.07\pm0.01$&$0.956\pm0.030\pm0.007$\\
\hline
Theory  &&&\\
LFQM   &$1.12$~\cite{tawfiq98} & $0.69$~\cite{tawfiq98} &
  $1.16$~\cite{tawfiq98}& $0.72$~\cite{tawfiq98}\\
RTQM &$1.78\pm0.33$~\cite{ivanov9899} &$1.26\pm0.17$~\cite{ivanov9899}&$2.11\pm0.29$~\cite{ivanov9899}
  & $1.01\pm0.15$~\cite{ivanov9899} \\
\hline
%&&&&\\
\multicolumn{1}{l}{}&\multicolumn{2}{c}
{$\Gamma(\Xi_c^{*\,+}\to\Xi_c^0\pi^++\Xi_c^+\pi^0)$}
&\multicolumn{2}{c}{$\Gamma(\Xi_c^{*\,0}\to\Xi_c^+\pi^-+\Xi_c^0\pi^0)$}\\
 &\multicolumn{2}{c}{[MeV]}
&\multicolumn{2}{c}{[MeV]}\\
\hline
\multicolumn{1}{l}{This work}&\multicolumn{2}{c}{$3.18\pm0.10\pm0.01$}&
\multicolumn{2}{c}{$3.03\pm0.10\pm0.01$}\\
\hline
\multicolumn{1}{l}{Experiment} & \multicolumn{2}{c}
{$<3.1$ (C.L.=90\%)~\cite{cleo96}}&\multicolumn{2}{c}{$<5.5$ 
(C.L.=90\%)~\cite{cleo95}}\\
\hline
Theory  &&&\\
CQM & \multicolumn{2}{c}{$<2.3\pm0.1$~\cite{rosner95}\ ,\ 1.191\,--\, 3.971~\cite{pirjol97}}  &
\multicolumn{2}{c}{$<2.3\pm0.1$~\cite{rosner95}\ ,\ 1.230\, --\, 4.074~\cite{pirjol97}}\\
% &  \multicolumn{2}{c}{1.191\,--\, 3.971~\cite{pirjol97}} & 
% \multicolumn{2}{c}{1.230\, --\, 4.074~\cite{pirjol97}}\\
 HHCPT & \multicolumn{2}{c}{$2.44\pm0.85$~\cite{cheng97}} &
  \multicolumn{2}{c}{$2.51\pm0.88$~\cite{cheng97}}\\
LFQM &  \multicolumn{2}{c}{1.81~\cite{tawfiq98}} &
 \multicolumn{2}{c}{1.88~\cite{tawfiq98}}\\
RTQM & \multicolumn{2}{c}{$3.04\pm0.50$~\cite{ivanov9899}} &
 \multicolumn{2}{c}{$3.12\pm0.33 $~\cite{ivanov9899}}\\
 \hline
\end{tabular}
%\end{center}
\caption{ \hspace{-.2cm} Decay widths for  
 $\Gamma(\Xi_c^{*\,+}\to\Xi_c^0\pi^+)$,
$\Gamma(\Xi_c^{*\,+}\to\Xi_c^+\pi^0)$,
$\Gamma(\Xi_c^{*\,0}\to\Xi_c^+\pi^-)$ and
$\Gamma(\Xi_c^{*\,0}\to\Xi_c^0\pi^0)$.
}
\label{tab:cas*cas}
\end{table}

Our results are stable against the use of  different
potentials with variations at the level of $6\sim8\%
$.
They are  in an overall good agreement with experimental data, 
in most cases in better agreement than predictions by other models.
%----------------------------------
\section{acknowledgments}
This research was supported by DGI and FEDER funds, under contracts
FIS2005-00810, BFM2003-00856 and  FPA2004-05616,  by the Junta de Andaluc\'\i a and
Junta de Castilla y Le\'on under contracts FQM0225 and
SA104/04, and it is part of the EU
integrated infrastructure initiative
Hadron Physics Project under contract number
RII3-CT-2004-506078. 
C. A. wishes to acknowledge a research contract with 
  Universidad de Granada. 
J.\ M. V.-V. acknowledges an E.P.I.F. contract with Universidad de Salamanca.

%---------------------------------

\end{document}